\documentclass[longauth]{aa} % for the long lists of affiliations 
\usepackage{graphicx}
\usepackage{txfonts}
\usepackage{color}
\usepackage{fixltx2e}
\usepackage{natbib}

\usepackage{hyperref}
\usepackage{breakurl}

\usepackage{lineno}
\newcommand*\patchAmsMathEnvironmentForLineno[1]{%
  \expandafter\let\csname old#1\expandafter\endcsname\csname #1\endcsname
  \expandafter\let\csname oldend#1\expandafter\endcsname\csname end#1\endcsname
  \renewenvironment{#1}%
     {\linenomath\csname old#1\endcsname}%
     {\csname oldend#1\endcsname\endlinenomath}}%
\newcommand*\patchBothAmsMathEnvironmentsForLineno[1]{%
  \patchAmsMathEnvironmentForLineno{#1}%
  \patchAmsMathEnvironmentForLineno{#1*}}%
\AtBeginDocument{%
\patchBothAmsMathEnvironmentsForLineno{equation}%
\patchBothAmsMathEnvironmentsForLineno{align}%
\patchBothAmsMathEnvironmentsForLineno{flalign}%
\patchBothAmsMathEnvironmentsForLineno{alignat}%
\patchBothAmsMathEnvironmentsForLineno{gather}%
\patchBothAmsMathEnvironmentsForLineno{multline}%
}

\newcommand{\fermilat}{\textit{Fermi}-LAT}
\newcommand{\swift}{\textit{Swift}-XRT}
\newcommand{\srcs}{QSO B0218+357}
\newcommand{\srclens}{B0218+357\,G}
\begin{document} 

\title{Detection of very high energy gamma-ray emission from the gravitationally-lensed blazar \srcs\ with the MAGIC telescopes} 
\titlerunning{Detection  of \srcs\ by the MAGIC telescopes.}
\subtitle{}

% authors 02.05.2016  Format AA
%
\author{
M.~L.~Ahnen\inst{1} \and
S.~Ansoldi\inst{2,}\inst{24} \and
L.~A.~Antonelli\inst{3} \and
P.~Antoranz\inst{4} \and
C.~Arcaro\inst{5} \and
A.~Babic\inst{6} \and
B.~Banerjee\inst{7} \and
P.~Bangale\inst{8} \and
U.~Barres de Almeida\inst{8,}\inst{25} \and
J.~A.~Barrio\inst{9}
% ...
}
\institute { ETH Zurich, CH-8093 Zurich, Switzerland
\and Universit\`a di Udine, and INFN Trieste, I-33100 Udine, Italy
\and INAF National Institute for Astrophysics, I-00136 Rome, Italy
\and Universit\`a  di Siena, and INFN Pisa, I-53100 Siena, Italy
\and Universit\`a di Padova and INFN, I-35131 Padova, Italy
\and Croatian MAGIC Consortium, Rudjer Boskovic Institute, University of Rijeka, University of Split and University of Zagreb, Croatia
\and Saha Institute of Nuclear Physics, 1/AF Bidhannagar, Salt Lake, Sector-1, Kolkata 700064, India
\and Max-Planck-Institut f\"ur Physik, D-80805 M\"unchen, Germany
\and Universidad Complutense, E-28040 Madrid, Spain
\and Inst. de Astrof\'isica de Canarias, E-38200 La Laguna, Tenerife, Spain; Universidad de La Laguna, Dpto. Astrof\'isica, E-38206 La Laguna, Tenerife, Spain
\and University of \L\'od\'z, PL-90236 Lodz, Poland
\and Deutsches Elektronen-Synchrotron (DESY), D-15738 Zeuthen, Germany
\and Institut de Fisica d'Altes Energies (IFAE), The Barcelona Institute of Science and Technology, Campus UAB, 08193 Bellaterra (Barcelona), Spain
\and Universit\"at W\"urzburg, D-97074 W\"urzburg, Germany
\and Institute for Space Sciences (CSIC/IEEC), E-08193 Barcelona, Spain
\and Technische Universit\"at Dortmund, D-44221 Dortmund, Germany
\and Finnish MAGIC Consortium, Tuorla Observatory, University of Turku and Astronomy Division, University of Oulu, Finland
\and Unitat de F\'isica de les Radiacions, Departament de F\'isica, and CERES-IEEC, Universitat Aut\`onoma de Barcelona, E-08193 Bellaterra, Spain
\and Universitat de Barcelona, ICC, IEEC-UB, E-08028 Barcelona, Spain
\and Japanese MAGIC Consortium, ICRR, The University of Tokyo, Department of Physics and Hakubi Center, Kyoto University, Tokai University, The University of Tokushima, KEK, Japan
\and Inst. for Nucl. Research and Nucl. Energy, BG-1784 Sofia, Bulgaria
\and Universit\`a di Pisa, and INFN Pisa, I-56126 Pisa, Italy
\and ICREA and Institute for Space Sciences (CSIC/IEEC), E-08193 Barcelona, Spain
\and also at the Department of Physics of Kyoto University, Japan
\and now at Centro Brasileiro de Pesquisas F\'isicas (CBPF/MCTI), R. Dr. Xavier Sigaud, 150 - Urca, Rio de Janeiro - RJ, 22290-180, Brazil
% ...
}

   \date{}

% \abstract{}{}{}{}{} 
% 5 {} token are mandatory
 
  \abstract
{
  \srcs\ is a gravitationally lensed blazar located at a redshift of 0.944.
  The gravitational lensing splits the emitted radiation into two components, spatially indistinguishable by gamma-ray instruments, but separated by a 10-12 day delay. 
  In July 2014, \srcs\ experienced a violent flare observed by the \fermilat\ and followed by the MAGIC telescopes.
}
{
  The spectral energy distribution of \srcs\ can give information on the energetics of $z \sim 1$ very high energy gamma-ray sources.
  Moreover the gamma-ray emission can also be used as a probe of the extragalactic background light at $z \sim 1$. 
}
{
  MAGIC performed observations of \srcs\ during the expected arrival time of the delayed component of the emission.
  The MAGIC and \fermilat\ observations were accompanied by quasi-simultaneous optical data from the KVA telescope and X-ray observations by \swift .
  We construct a multiwavelength spectral energy distribution of \srcs\ and use it to model the source.
  The GeV and sub-TeV data, obtained by \fermilat\ and MAGIC, are used to set constraints on the extragalactic background light. 
}
{
  Very high energy gamma-ray emission was detected from the direction of \srcs\ by the MAGIC telescopes during the expected time of arrival of the trailing component of the flare, making it the farthest very high energy gamma-ray sources detected to date. 
  The observed emission spans the energy range from 65 to 175 GeV.
  The combined MAGIC and \fermilat\ spectral energy distribution of \srcs\ is consistent with current extragalactic background light models.
  The broad band emission can be modeled in the framework of a two zone external Compton scenario, where the GeV emission comes from an emission region in the jet, located outside the broad line region.
}
{}


   \maketitle
%
%________________________________________________________________

\section{Introduction}
Even though there are already over 60 blazars detected in the very high energy (VHE, $\gtrsim100\,$GeV) range, most of them are relatively close-by sources with redshift $z\lesssim0.5$.
Until mid 2014, the farthest sources observed in this energy range were 3C\,279 ($z=0.536$, \citealp{al08}), KUV\,00311-1938 ($z>0.506$, \citealp{be12}) and PKS1424+240 ($z=0.601$, \citealp{acc10}).
In the last two years the MAGIC (Major Atmospheric Gamma Imaging Cherenkov) telescopes discovered VHE gamma-ray emission from \srcs\ at $z=0.944$ \citep{atel6349} and afterwards PKS1441+25 at $z=0.940$ \citep{ah15} almost doubling the boundaries of the known gamma-ray universe. 
Observations of distant sources in VHE gamma-rays are difficult due to strong absorption in the extragalactic background light (EBL, see e.g. \citealp{gs66}).
At a redshift of $\sim 1$ it results in a cut-off at energies\footnote{Unless specified otherwise, the energies are given in the Earth's frame of reference} $\sim 100\,$GeV.
Such energies are at the lower edge of the sensitivity range of the current generation of Imaging Atmospheric Cherenkov Telescopes (IACTs), making such observations challenging.
To maximize the chance of detection, the observations are often triggered by a high state observed in lower energy ranges. 
In particular, \fermilat\ (Large Area Telescope) scanning the whole sky every 3 hours provides alerts on sources with high fluxes and information about the spectral shape of the emission in the GeV range. 

\srcs , also known as S3\,0218+35, is classified as a flat spectrum radio quasar (FSRQ, \citealp{ac11}).
The classification is based on the optical spectrum \citep{co03}. 
It is located at a redshift of $z_s=0.944\pm0.002$ \citep{co03}. %\citep{ab10b}. %\citep{li12}.
One of the five features from which \citet{co03} derived the redshift was confirmed by \citep{la15}.
The object is gravitationally lensed by the face-on spiral galaxy \srclens\ located at a redshift of $z_l=0.68466\pm0.00004$ \citep{cry93}.
Strong gravitational lensing forms multiple images of the source \citep[see e.g.][]{ko04}.
The flux magnification of an image is the ratio of the number of photons gravitationally deflected into a small solid angle centered on the observer to the number of photons emitted by the source in such a solid angle.
The 22.4 GHz VLA radio image shows two distinct components with an angular separation of only 335\,mas and an Einstein ring of a similar size \citep{od92}.
Observations of variability of the two radio components led to a measurement of a delay of 10-12 days between the leading and trailing images \citep{co96,bi99, co00, em11}. 
In the radio image, the leading component (also dubbed `image A' in literature) is located to the west from the trailing component (image B).
The delayed component had a 3.57-3.73 times weaker flux \citep{bi99}.
However, the observed ratio of magnification varies with the radio frequency \citep{mi06}, presumably due to free-free absorption in the lensing galaxy \citep{mi07}.
In the optical range the leading image is strongly absorbed \citep{fa99}.

In 2012 \srcs\ went through a series of outbursts registered by the \fermilat\ \citep{ch14}.
Even though \fermilat\ does not have the necessary angular resolution to disentangle the two emission components, the statistical analysis of the light curve auto-correlation function led to a measurement of a time delay of $11.46\pm0.16$ days. 
Interestingly the average magnification factor, contrary to radio measurements, was estimated to be $\sim1$.
Changes in the observed GeV magnification ratio were interpreted as microlensing effects on individual stars in the lensing galaxy \citep{vn15}. 
Microlensing on larger scale structures has been considered as well \citep{sb16}.
The radio follow-up observations of \srcs\ after the 2012 gamma-ray flare did not reveal any correlation between the two bands \citep{sp16}.

Another flaring state of \srcs\ was observed by \fermilat\ on 2014 July 13 and 14 \citep{atel6316}. 
Contrary to the results for the 2012 flaring period, in this case the ratio of the leading to delayed GeV emission was at least 4 \citep{bu15}.
The 2014 flare triggered follow-up observations by the MAGIC telescopes, which in turn led to the discovery of VHE gamma-ray emission from \srcs\ \citep{atel6349}.

In this work we present the results of the observations by the MAGIC telescopes and supporting multiwavelength instruments of the \srcs\ during the flaring state in July 2014.
In Section~\ref{sec:inst} we describe the instruments taking part in those observations and the data reduction. 
The effect of the lensing galaxy on the observed emission is discussed in Section~\ref{sec:lens}.
Section~\ref{sec:results} is devoted to the results of the observations.
In Section~\ref{sec:sed} we model the spectral energy distribution (SED) of the source.
We use the \fermilat\ and MAGIC observations to discuss constraints on the EBL in Section~\ref{sec:ebl}.

%__________________________________________________________________
\section{Instruments, observations and analysis}\label{sec:inst}
The VHE gamma-ray observations of \srcs\ during the flaring state in July 2014 were performed with the MAGIC telescopes. 
The source was also monitored in GeV energies by \fermilat , in X-ray by \swift\ and in optical by KVA.

\subsection{MAGIC}
MAGIC is a system of two 17\,m Cherenkov telescopes located in the Canary Island of La Palma at a height of 2200 m a.s.l.
The MAGIC telescopes combine large mirror area, allowing us to observe gamma rays with energies as low as $\sim 50\,$GeV, with the stereoscopic technique providing strong hadronic background rejection, and hence good sensitivity at low energies.
This makes them an excellent instrument for observations of distant FSRQs.
In summer 2012 MAGIC finished a major upgrade \citep{al16a} greatly enhancing the performance of the instrument \citep{al16b}.
The sensitivity\footnote{defined as the flux of the source with a Crab-like spectral shape that gives a gamma-ray excess with a significance of 5$\sigma$} of the MAGIC telescopes achieved in the energy range $\gtrsim100\,$GeV is at the level of $1.45\%$ of Crab Nebula flux in 50\,h of observations.
The angular resolution of MAGIC is of the order of $0.09^\circ$, i.e. insufficient for spatially resolving the emission from the two lensed image components of \srcs .

The telescopes could not immediately follow the flare alert published by \fermilat\ in mid July 2014 from \srcs\ as it occured during the full Moon time (the Cherenkov light from low energy gamma-ray showers would be hidden in the much larger noise from the scattered moonlight).
The MAGIC observations started 10 days later, with the aim of studing the possible emission during the delayed flare component. 
The observations were performed during 14 consecutive nights from the 23 of July (MJD=56861, two nights before the expected delayed emission) to the 5 of August 2014 (MJD=56874).
The total exposure time was 12.8\,h and the source was observed at an intermediate zenith angle ($20^\circ-43^\circ$).
The data reduction (stereo reconstruction, gamma/hadron separation and estimation of the energy and arrival direction of the primary particle) was performed using the standard analysis chain of MAGIC \citep{magic_mars, al16b}.
The sky position of \srcs , contrary to the Crab Nebula used to estimate the MAGIC telescopes performance in \cite{al16b}, is not projected against the Milky Way optical background. 
A $30\%$ smaller night sky background registered by the MAGIC telescopes for \srcs\ allowed us to apply image cleaning thresholds lower by $15\%$ with respect to the ones used in the standard analysis presented in \citet{al16b}.
For the zenith angle range in which the observations were performed this resulted in the analysis energy threshold of about $85$\,GeV (measured as the peak of the Monte Carlo (MC) energy distribution\footnote{Note that it is also possible to reconstruct the flux slightly below such a defined threshold.} for a source with the spectral shape of \srcs )
The lower image cleaning thresholds were validated by applying the same procedure to the so-called pedestal events, i.e. events which contain only the light of the night sky and electronic noise.
An acceptable fraction of about $10\%$ of such images survived the image cleaning.
The analysis was performed using a dedicated set of MC simulations of gamma rays with the night sky background and the trigger parameters tuned to reproduce as accurately as possible the actual observation conditions.

\subsection{\fermilat }
\fermilat\ is a pair-conversion telescope optimized for energies from 20 MeV to greater than 300 GeV \citep{at09}.
Generally, \fermilat\ is operated in scanning mode, providing coverage of the full sky every three hours. 
Starting on December 2013 and until December 2014, a new observing strategy that emphasized coverage of the Galactic center region was adopted. 
\srcs\ data presented in this paper were obtained during this time interval. 
As a consequence, the coverage on the blazar position was on average a factor of 0.6 of the maximum one. 
Additionally, at the time of the expected delayed emission, Fermi performed a ToO (Target of Opportunity) observation on \srcs\ to enhance exposure toward the source position. 
The ToO lasted approximately 2.7 days (2014-07-24 00:30:01 UTC to 2014-07-26 18:24:00 UTC, MJD$=56862.02-56864.77$).

\fermilat\ data were extracted from a circular region of interest (ROI) of $15^{\circ}$ radius centered at the \srcs\ radio position, ${\rm~R.A.}=35^\circ.27279$, ${\rm~Decl.}=35^\circ.93715$ \citep[J2000;][]{pat92}.
The analysis was done in the energy range $0.1 - 300$ GeV using the standard \textit{Fermi Science Tools} (version {\tt v9r34p1}) in combination with the {\tt P7REP\_SOURCE\_V15} LAT Instrument Response Functions.
For obtaining the light curve, data collected between MJD=56849--56875 (2014 July 11 -- 2014 August 6) were used. 
For the spectral analysis only data spanning the two days during the flare observed by MAGIC (MJD=56863.125 - 56864.5) were used.
We applied the {\tt gtmktime} filter ($\#$3) cuts to the LAT data following the FSSC recommendations\footnote{\burl{http://fermi.gsfc.nasa.gov/ssc/data/analysis/documentation/Cicerone/Cicerone\_Likelihood/Exposure.html}}. 
According to this prescription, time intervals when the LAT boresight was rocked with respect to the local zenith by more than $52^{\circ}$ (usually for calibration purposes or to point at specific sources) and events with zenith angle $>100^{\circ}$  were excluded to limit the contamination from Earth limb photons. 

The spectral model of the region included all sources located within the ROI with the spectral shapes and the initial parameters for the modeling set to those reported in the third \fermilat\ source catalog  \citep[3FGL,][]{3fgl} as well as the isotropic  ({\tt iso\_source\_v05.txt}) and  Galactic diffuse  ({\tt gll\_iem\_v05.fit}) components\footnote{\burl{http://fermi.gsfc.nasa.gov/ssc/data/access/lat/BackgroundModels.html}}.
For generating the light curve the source of interest was modeled with a power-law spectral shape with normalization and index free to vary.
To access the detection significance we used the Test Statistic (TS) value.
It is defined as TS = $-2\log (L_0 /  L)$,  where $L_0$ is the maximum likelihood value for a model without an additional source (the `null hypothesis') and $L$ is the maximum likelihood value for a model with the additional source at the specified location.
The TS quantifies the probability of having a point gamma-ray source at the location specified and corresponds roughly to the square of the standard deviation assuming one degree of freedom \citep{mat96}.
As in our analysis the second model had two more degrees of freedom (i.e. normalization and index were left free), therefore TS=9 (25) corresponds to significance of $\sim$2.5 (4.6) $\sigma$, respectively.
During the analyzed period, \srcs\ was not always significantly detected. Flux upper limits at the 95\% confidence level were calculated for each interval where the source TS was $<9$.

\subsection{\textit{Swift}}
\srcs\ was observed by the \textit{Swift} satellite during 10 epochs, each with an exposure of about 4.5 ks.
The observations did first follow the original alert of enhanced activity in GeV gamma rays, and then were resumed at the expected time of arrival of the delayed component. 
The data were reduced with the \texttt{HEASoft} package version 6.17.
The \textit{Swift} X-ray Telescope (XRT, \citealp{bu05}) is a CCD imaging spectrometer, sensitive in the 0.2-10 keV band. 
We reduced the data using the calibration files available in the version 20140709 of the \textit{Swift}-XRT CALDB.
We run the task \texttt{xrtpipeline} with standard screening criteria on the observations performed in pointing mode. 
Observations were done in Photon Counting (PC) mode with count rates about 0.02 counts/s. 
The weak X-ray emission compelled us to merge different epochs to create a good quality spectrum (see Section~\ref{sec:swift2}). 
We combined different event files with the task \texttt{xselect} summing the corresponding exposure maps with the task \texttt{image}. 
The merged source and background counts were extracted with the task \texttt{xrtproducts} from a circular region of 35" for the source and 120" for the background. 
We grouped each spectrum with the corresponding background, redistribution matrix (rmf), and ancillary response files (arf) with the task \texttt{grppha}, setting a binning of at least 20 counts for each spectral channel in order to use the $\chi^2$ statistics. 
The spectra were analyzed with \texttt{Xspec} version 12.8.1. 
We adopted a Galactic absorption of $N_{\rm H} = 5.6 \times 10^{20}\,\mathrm{cm}^{-2}$ from the LAB survey \citep{ka05}.

Simultaneous observations by the Ultraviolet Optical Telescope (UVOT, \citealp{ro05}), on board of {\em Swift}, did not result in a significant detection of the emission from the source in the UV range.

\subsection{KVA}

The optical R-band observations were done using the 35 cm Celestron telescope attached to the KVA 60 cm telescope (La Palma, Canary islands, Spain). 
The observations started on  2014, July 24 (MJD=56862.2) and continued on almost nightly basis until 2014, August 5 (MJD=56874.2). 
Further follow-up observations were performed in August and September.
The data have been analyzed using the semi-automatic pipeline developed at the Tuorla Observatory (Nilsson et al. 2016, in prep.). 
The magnitudes are measured using differential photometry.
We performed absolute calibration of the optical fluxes using stars with known magnitudes present in the field of view of the instrument during observations of all targets of a given night (see Table 3 of \citet{ni07} and references therein). 
\srcs\ is rather faint in the optical range (about 19 mag) and the telescope is relatively small, therefore several images from the same night were combined for the measurement of the average flux.
For the spectral analysis the optical flux was deabsorbed using a galactic extinction of $A_{\rm R}=0.15$ \citep{sf11}.

\section{Influence of the lensing galaxy} \label{sec:lens}
The interpretation of the \srcs\ observations is not trivial due to the influence of the lensing galaxy.
The lensing introduces a (de-)magnification factor to the observed flux with respect to the intrinsic emission, due to both lensing by the galaxy itself, which is additionally causing the time delay between the images, and microlensing by individual stars in the lens galaxy \citep{vn15}. 
Variability of the flux magnification caused by microlensing is larger (and thus can more significantly affect the measured light curve) for smaller emission regions in the source.
Using a simple Singular Isothermal Sphere model (SIS, see e.g. \citealp{ko04}) we roughly estimate the absolute magnification of the leading and trailing images. 
A rigorous lens modelling performed by \citet{ba15} yielded a model consistent with SIS. 

The ratio between the observed angular distances to the lens of the leading and trailing radio images of the source has been measured to be $\sim 4$ \citep{yo05}.
In the framework of SIS model this results in the individual magnifications of the two images to be $\mu_{\rm leading}\approx 2.7$, $\mu_{\rm trailing}\approx 0.67$.

Using the flux ratio between the two images measured in the radio frequency range, the same model allows us to also compute the absolute magnifications independently.
With the value of $\mu_{\rm leading}/\mu_{\rm trailing}\approx3.6$ \citep{bi99} we obtain very similar results $\mu_{\rm leading}\approx 2.8$, $\mu_{\rm trailing}\approx 0.77$. 
Averaging both methods we assume $\mu_{\rm leading}\approx 2.7$, $\mu_{\rm trailing}\approx 0.7$ in the further calculations.
The radio emission in blazars is believed to originate from regions much larger than the ones involved in gamma-ray production.
Therefore, the values given above for the individual magnifications of images are not affected by possible microlensing on individual stars of \srclens .

On the other hand the microlensing can significantly modify the fluxes observed in the HE and VHE energy ranges  \citep{nvm15, vn15}. 
The flux magnification due to microlensing depends on the size of the emission region, which might vary with the energy e.g. due to cooling effects.
Thus, it might modify the observed spectrum and this, in principle, can affect the EBL constraints and source modelling.
However, during the 2014 flaring period the magnification ratio observed in \fermilat\ was comparable to, or larger than, the radio one.
This suggests that the microlensing, if present, might have a bigger effect on the leading rather than on the trailing image, which was observed by MAGIC. 
Namely, if a microlensing event amplified the observed emission during the delayed flare with a given magnification of $\mu_{\rm star,trailing}$, the leading flare must have been also amplified with even larger magnification $\mu_{\rm star,leading}\gtrsim \mu_{\rm star,trailing}$ by an independent microlensing event to keep the observed ratio of fluxes.
Assuming that a probability that the flux of the trailing image is magnified with a factor of $\mu_{\rm star,trailing}$ is $p_{\rm trailing}$, the probability that both images are independently magnified resulting in the observed flux ratio is much smaller, roughly $\lesssim p_{\rm trailing}^2$.

Absorption in the lensing galaxy can also affect the observed fluxes at different energies.
\cite{fa99} interpreted the different reddening of the two images of \srcs\ as an additional absorption of the leading image with the differential extinction $\Delta E(B-V)=0.90\pm0.14$. 
In fact the absorption is so strong that it inverts the brightness ratio of the two images in the optical range, making the trailing image brighter. 
Also, in the leading image, the H$_2$ column density was estimated at the level of  $0.5-5\times 10^{22}\,\mathrm{[cm^{-2}]}$ by an observation of a molecular absorption \citep{mr96}. 
In addition the dependence of the radio flux ratio on the frequency could also stem from free-free absorption \citep{mi07}.
No absorption has been measured for the trailing image. 
Observations of 21 cm absorption feature in B0218+357 points to an HI column density of $10^{21}(T_s/100\,\mathrm{K})/(f/0.4)\,\mathrm{[cm^{-2}]}$, where $T_S$ is the spin temperature and $f$ is the fraction of the flux density obscured by HI \citep{cry93}.
The absorption of sub-TeV emission is expected to be negligible in the lensing galaxies \citep{bbs14}.

\section{Results}\label{sec:results}
In this section we discuss the spectral and temporal characteristics of the \srcs\ emission obtained in different energy bands.

\subsection{MAGIC}
\begin{figure}
\includegraphics[width=0.49\textwidth]{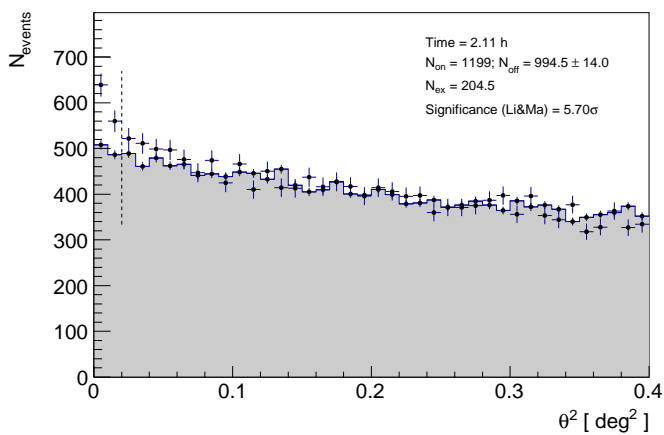}
\caption{Distribution of the squared angular distance, $\theta^2$, between the reconstructed source position and the nominal source position (points) or the background estimation position (shaded area). 
The vertical dashed line shows the value of $\theta^2$ up to which the number of excess events and significance are computed. 
}\label{fig:th2}
\end{figure}

The VHE gamma-ray emission was detected on the nights of 25 and 26 of July 2014 (MJD=56863.2 and 56864.2 respectively), during the expected arrival time of the delayed component of the flare registered by \fermilat .
The detection cuts were optimized to provide the best sensitivity in the 60-100 GeV estimated energy range (see \citealp{al16b}).  
The total observation time during those 2 nights of 2.11\,hr yielded a statistical significance, computed according to \cite{lm83}, Eq.\,17, of $5.7 \sigma$ (see Fig.\,\ref{fig:th2}). 

\begin{figure}
\includegraphics[width=0.49\textwidth]{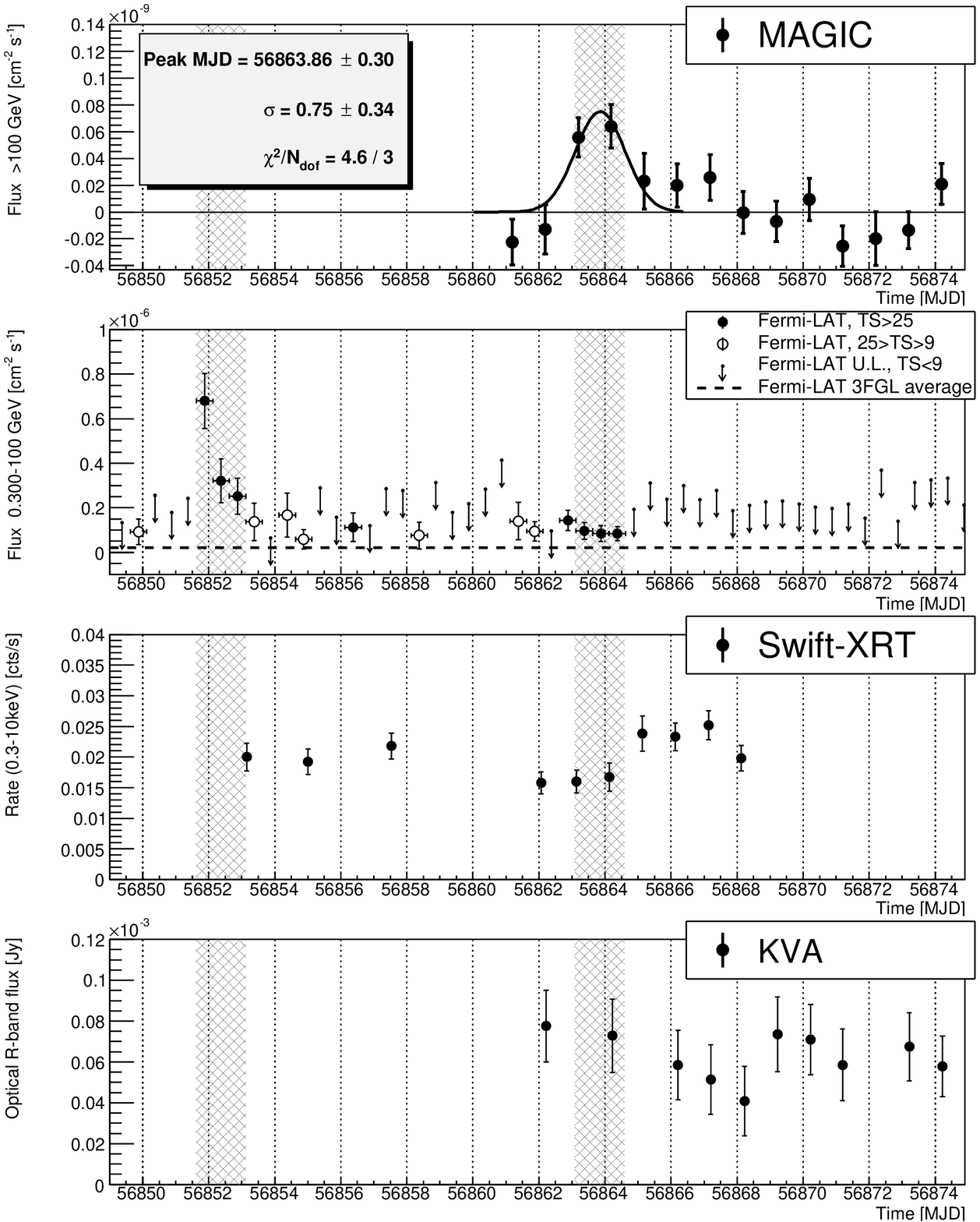}
\caption{
Light curve of \srcs\ during the flaring state in July/August 2014. 
Top panel: MAGIC (points) above 100\, GeV and a Gaussian fit to the peak position (thick solid line). 
Second panel from the top: \fermilat\ above 0.3\,GeV with the average flux from the 3rd Fermi Catalog \citep{3fgl} marked with a dashed line.
Notice that, during the days where the trailing emission was expected \fermilat\ was in pointing mode allowing the significant detection of lower flux levels.
Third panel from the top: \swift\ count rate in the 0.3-10\,keV range.
Bottom panel: KVA in R band (not corrected for the contribution of host/lens galaxies and the Galactic extinction).  
The two shaded regions are separated by 11.46 days.
}\label{fig:lc}
\end{figure}

The light curve above 100\,GeV is shown in Fig.\,\ref{fig:lc}.
A fit with a Gaussian function gives the peak position at MJD=$56863.86 \pm 0.30_{\rm stat}$ and a standard deviation of $0.75 \pm 0.34_{\rm stat}$ days. 
The corresponding fit probability\footnote{The fit probability throughout the paper is defined as the probability that the observed $\chi^2$ of points distributed along the used model shape exceeds by chance the value of $\chi^2$ obtained from fitting the data points with this model.} is 21\%. 
The two flaring nights give a mean flux above 100\,GeV of $(5.8 \pm 1.6_{\rm stat} \pm 2.4_{\rm syst})\times 10^{-11}\mathrm{cm^{-2}\,s^{-1}}$.
The relatively large systematic error is mainly due to the 15\% uncertainty in the energy scale.

\begin{figure}
\includegraphics[width=0.49\textwidth]{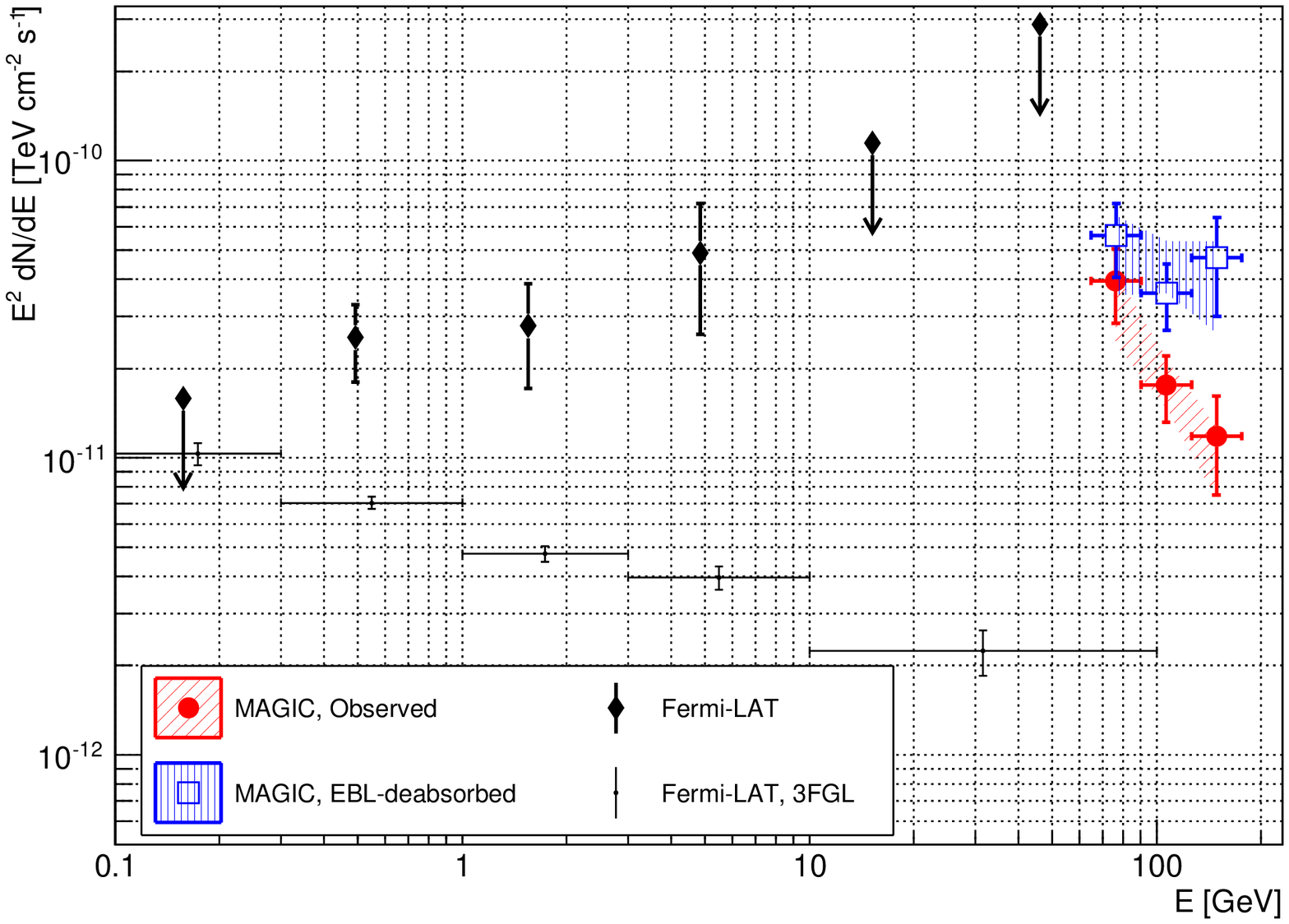}
\caption{
Gamma-ray SED of \srcs\ as observed during the two flaring nights, 25 and 26 of July, by MAGIC (red filled circles) and after deabsorption in EBL according to the \cite{do11} model (blue open squares).
The shaded regions show the 1 standard deviation of the power-law fit to the MAGIC data.
Black diamonds show the \fermilat\ spectrum from the same time period.
Black points show the average emission of \srcs\ in the 3FGL catalog \citep{3fgl}. 
}\label{fig:magicsed}
\end{figure}

The SED obtained from the two nights 25 and 26 of July (MJD=$56863.2$ and $56864.2$) is presented in Fig.\,\ref{fig:magicsed}.
The reconstructed spectrum spans the energy range $65-175$\,GeV and can be described as a power law:
\begin{equation}
dN/dE = f_0  \times \left(E/100\,\mathrm{GeV}\right)^{-\gamma}\ 
\end{equation}
with the fit probability of 47\%.
The parameters obtained are: $f_0=(2.0 \pm 0.4_{\rm stat} \pm 0.9_{\rm syst}) \times 10^{-9} \mathrm{cm^{-2} s^{-1} TeV^{-1}}$ and $\gamma=3.80\pm0.61_{\rm stat} \pm 0.20_{\rm syst}$.
The quoted systematic uncertainty on the spectral index takes into account also the small background estimation uncertainty for a weak low-energy source (see Eq.~3 of \citealp{al16b}).
As the redshift of the source is close to 1 the spectrum is severely affected by the absorption of VHE gamma-rays in the EBL. 
Correcting the observed spectrum for such absorption modelled according to \cite{do11}, we obtain an intrinsic spectral index of $2.35\pm0.75_{\rm stat} \pm 0.20_{\rm syst}$. 
The corresponding normalization of the emission at 100\,GeV is $(4.6 \pm 0.8_{\rm stat}  \pm 2.1_{\rm stat}) \times 10^{-9}\ \mathrm{cm^{-2} s^{-1} TeV^{-1}}$.
The spectral points are obtained using the Bertero unfolding method, while the fit parameters are obtained using the so-called forward unfolding \citep{al07}.   

\subsection{\fermilat\ }
The GeV light curve of \srcs\ is shown in the second panel of Fig.\,\ref{fig:lc}.
We used a minimum energy 0.3\,GeV in the light curve (instead of 0.1\,GeV) in order to increase the signal to noise ratio in the flux measurements: the spectrum of this source during this flaring episode is very hard (see below), while the diffuse backgrounds fall with energy with an index of $>2.4$, and the PSF of LAT at 0.1\,GeV is about twice larger than that at 0.3\,GeV.
Significant GeV gamma-ray emission was detected by \fermilat\ both during the leading flare and during the expected arrival time of the delayed emission (TS of 615 and 129 respectively). 
The spectrum contemporaneous to the MAGIC detection, derived between MJD=56863.07 and 56864.85, can be described by a power-law function with slope $\gamma=1.6\pm0.1$.
The corresponding flux above 0.1\,GeV is $F_{>0.1\mathrm{GeV}}=(1.7\pm0.4)\times 10^{-7} \mathrm{cm^{-2}\,s^{-1}}$.
For comparison, the leading flare was marginally harder, $\gamma=1.35\pm0.09$, with $\sim 4$ times higher flux $F_{>0.1\mathrm{GeV}}=(6.7\pm1.0)\times 10^{-7} \mathrm{cm^{-2}\,s^{-1}}$
The spectral index measured by \fermilat\ during the 2014 outburst, is much harder than $\gamma \sim 2.3$ during both the flaring period of 2012 \citep{ch14} and the average state of this source reported in Third \fermilat\ Catalog \citep{3fgl}.

\subsection{\swift}\label{sec:swift2}
In the third panel of Fig.\,\ref{fig:lc} we present the X-ray light curve of \srcs .
The whole observed light curve shows only a small hint of variability. 
A constant fit gives $\chi^2/N_{\rm dof}=21.3/9$, corresponding to the probability of 1.1\%.
The source did not show an enhanced flux in the X-ray range during the trailing gamma-ray flare. 
The average count rate from the two observations during the enhanced gamma-ray flux results in $(79.2\pm7.7)\%$ of the rate averaged from the remaining 8 pointings. 
The rate obtained in the 0.3-10\,keV energy range is similar to the one obtained during the 2012 flaring period ($0.027\pm0.003$, \citealp{do12}).

As the source is a weak X-ray emitter and the observed variability is not very strong we have combined all the pointings for the spectral modelling of the source.
Moreover, the lack of strong variability also implies that the observed emission is the sum of the two images of the source, with at least one of them affected by the hydrogen absorption. 
In order to provide higher accuracy per spectral point, we rebin the spectrum to 50 events per bin.

\begin{figure}
\includegraphics[width=0.49\textwidth]{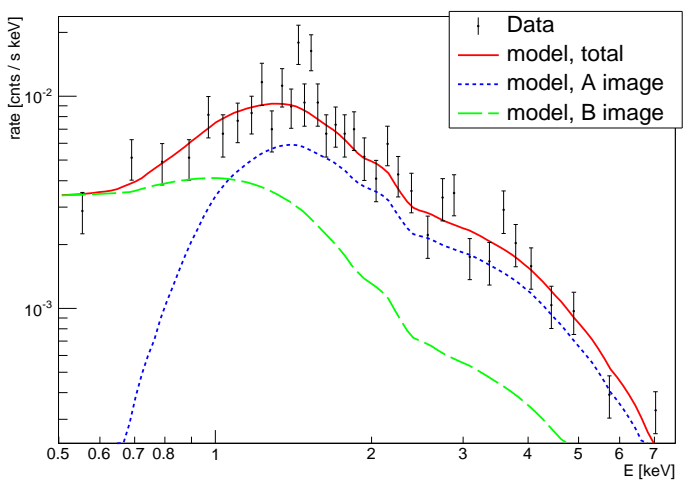}
\caption{
Energy-binned counts observed by \swift\ from the direction of \srcs\ (data points).
The emission is modeled as a sum (solid red line) of two power-law components with the same spectral index. 
The first component (A image) is magnified by a factor 2.7  with an additional strong hydrogen absorption at the lens (dotted blue line). 
The second component (B image) is intrinsically weaker (magnification factor 0.7), but not absorbed at the lens (dashed green line). 
}\label{fig:swiftspectrum}
\end{figure}

We model the X-ray spectrum as a sum of two power-law components, with the same intrinsic normalization and spectral slope, but magnifications fixed to 2.7 and 0.7 respectively (see Section~\ref{sec:lens}). 
Following the detection by \cite{mr96} of the molecular absorption line in the brighter image, we include hydrogen absorption at the redshift of the lens in the first (brighter) component. 
However, due to large uncertainty in the hydrogen column density, we leave it as a free parameter of the model.
With such assumptions, X-ray intrinsic spectrum can be well described ($\chi^2/N_{\rm dof} = 42.2/34$) by a simple power law  (see Fig.\,\ref{fig:swiftspectrum}).

We obtained the following spectral fit parameters: 
\begin{equation}
\frac{dN}{dE} = (2.69\pm0.29)\times 10^{-4} \left(\frac{E}{\mathrm{keV}}\right)^{-1.90\pm0.08} \mathrm{[keV^{-1} cm^{-2} s^{-1}]},
\end{equation}
where the reported uncertainties are statistical only. 
The corresponding column density $(2.4\pm0.5)\times 10^{22}\ \mathrm{at.\,cm^{-2}}$ is within the bounds given by the radio measurement of \cite{mr96}.

The X-ray spectrum can be alternatively described by a simpler model, considering only absorption of the total emission (i.e. same absorption is affecting both images). 
The resulting spectrum is then slightly harder, with an index of $1.59\pm0.10$.
The corresponding effective hydrogen column density is smaller, $(0.57\pm0.17)\times 10^{22}\ \mathrm{at.\,cm^{-2}}$.
The fit probability is however worse in this case, with $\chi^2/N_{\rm dof} = 54.7/34$.
Therefore, for the SED modelling (see Section~\ref{sec:sed}) we use the spectrum obtained using the assumption that the absorption affects only the leading image.

\subsection{KVA}
The bottom panel of Fig.\,\ref{fig:lc} shows the optical light curve of \srcs\ in the R band.
In all of our observations the source was fainter than 19 magnitudes. 
The resulting error bars for the flux points were therefore relatively large and no significant variability was detected. 
We estimated the observed galaxy flux within our measurement aperture (5.0 arcsec radius) and taking into account that the calibration is made through an aperture of the same size. 
Using the data in \citet{le00} a lens galaxy flux of $F_{\rm galaxy} = 13\,\mu$Jy was derived.
The resulting flux (corrected for both the Galactic absorption and the galaxy contribution) for the observation during the flare is then $70\pm20\,\mu$Jy.

\section{Modelling of the broadband emission}\label{sec:sed}

\begin{figure}
\begin{minipage}{0.49\textwidth}
\includegraphics[width=\textwidth]{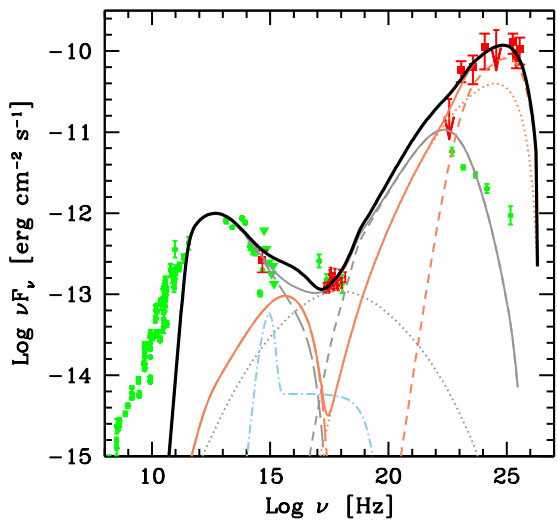}
\caption[Broadband SED]{
Broadband SED of \srcs\ modeled with a two-zone model. 
The reconstructed fluxes (red squares) are corrected for different magnifications in different energy ranges (see the text). 
Historical measurements (ASDC\footnote{see \burl{http://www.asdc.asi.it/}}) are shown with green circles and triangles (flux upper limit).
Gray curves depict the emission from the region located within the BLR, while orange curves refer to the region located beyond the BLR. 
Long dashed curves show the  synchrotron radiation, dotted the SSC emission and short dashed the external Compton emission.
Dashed-dotted light blue line represents the accretion disk emission and its X-ray  corona.
The solid black line shows the sum of the non-thermal emission from both regions.
}\label{fig:mwlsed}
\end{minipage}
\end{figure}

In order to model the broadband emission spectrum of \srcs\ we need to determine the magnification factors affecting different energy ranges and correct for them. 
As no strong variability was seen in either the optical or the X-ray range we can assume that the observed emission in those energy ranges is the sum of both lensed images.
However, the optical leading image is strongly absorbed \citep{fa99}, thus the total magnification in the optical range is close to $\mu_{\rm trailing}$.
On the other hand in the X-ray range the emission $\gtrsim2\,$keV is not strongly absorbed in either of the two images.
We correct the absorption of softer X-rays in the analysis (see Section~\ref{sec:swift2}).
Therefore we assume that the magnification in the X-ray energy range is $\mu_{\rm leading}+\mu_{\rm trailing}\approx3.4$.
The strong variability in the GeV and sub-TeV gamma-ray range and the much harder GeV spectrum during the MAGIC observations point to the magnification in the GeV energy range at this time to be close to $\mu_{\rm trailing}$.
The broad-band SED of \srcs\ demagnified according to the numbers derived above and corrected for the X-ray and gamma-ray absorption is shown in Fig.\,\ref{fig:mwlsed}.
In green we report historical data, obtained from ASDC (ASI Science Data Center, see \burl{http://www.asdc.asi.it/}), tracking particularly well the low energy component.
These historical data are the sum of the emission of the source passing through both of its images, however especially in optical and UV range are affected by strong absorption in the leading component. 
In order to derive the intrinsic flux of the source we apply the following correction factor to the flux $1/(\mu_{\rm trailing} + \mu_{\rm leading}\times T_A(f))$, where $T_A(f)$ is the frequency-dependent fraction of the leading image flux surviving the attenuation.
In order to estimate $T_A(f)$ we use the differential extinction of the leading image, $\Delta E(B-V)=0.90\pm0.14$ \citep{fa99}, to scale the dust extinction curve of the Milky Way \citep{pe92,xu16}, taking also into account the redshift of the lens. 
In the X-ray range the $T_A(f)$ shape was determined from the hydrogen column density obtained in Section~\ref{sec:swift2}.

The SED is dominated by the emission at GeV -- sub-TeV  energies,  which is relatively common in flaring FSRQs (see e.g. \citealp{al14, pa14}). 
Although the corrections for lensing are uncertain, the intrinsic GeV spectrum appears to be hard for this flaring state.

Interestingly, the gamma-ray flare seen by MAGIC was not accompanied by a similar increase in either optical or in X-ray flux.
This is unusual for FSRQs, where a correlation is often seen.
Comparison of the optical data to the archival measurements, shows that there was no large change in the position of the low energy peak during the high-energy flare.
The high energy peak position however moved from the sub-GeV range in low state to tens of GeV during the flare. 

\subsection{One zone leptonic models}
As the two peaks have a large separation in frequency the resulting spectrum cannot be easily explained in the framework of one-zone models. 
As detailed in \citet{tg14} in such a case one-zone models inevitably require quite large Doppler factors and very low magnetic fields. 
For the specific case of \srcs\, in order to explain the high energy peak photons with frequency $\nu_{C}=10^{25} \nu_{C,25}\mathrm{Hz} \simeq 10^{25}$ Hz, the Lorentz factors of the electrons emitting at the peak would need to reach (or exceed in case the scattering is in the Thomson regime) $8\times10^4 \nu_{C,25} \delta^{-1} (1+z_s)$, where $\delta $ is the Doppler factor of the emitting region.
Since the same electrons are responsible also for the synchrotron radiation of the low energy peak, a very low value of the magnetic field is required:
$B= 5.6\times 10^{-5} \nu_{s,12} \nu_{C,25}^{-2} \delta^{-1} (1+z_s) [\mathrm{G}]$, where $\nu_{s,12}$ is the synchrotron peak frequency in units of $10^{12}$ Hz. 
If the high-energy component is interpreted as synchrotron-self-Compton (SSC) emission, the ratio of the high-energy peak luminosity to the synchrotron luminosity has to be equal to the ratio between the synchrotron photon energy density and the magnetic field energy density. 
This condition, coupled to the value of the magnetic field derived above, allows us to derive the required Doppler factor (see \citealp{tg14} for details):
$\delta \gtrsim 35 L_{s,46}^{1/8} \nu_{s,12}^{-1/4} \nu_{C,25}^{1/2} \Delta t_{1d}$, where $L_{s,46}$ is the synchrotron luminosity (measured in the units of $10^{46}\,\mathrm{erg\,s^{-1}}$) and $\Delta t_{1d}$ is the variability timescale in units of days.
With such a large value of the Doppler factor it is rather unlikely that the radiation energy density in the jet frame is dominated by the synchrotron one. 
Instead, as usually assumed in the modeling of a FSRQ (e.g. \citealp{si94}), it is most plausible that the high-energy component is produced by the scattering of external photons (from the broad line region, BLR, the disk or the molecular torus). 

In the case of such an external Compton (EC) scenario, some constraints can be derived considering that the SSC emission, expected now to peak in the X-ray band, cannot have a flux exceeding the  value fixed by the XRT data. 
Similarly to the discussion above for the SSC case, we can thus derive a constraint on the Doppler factor:
$\delta \gtrsim 75 L_{s,46}^{1/8} \nu_{s,12}^{-1/4} \nu_{C,25}^{1/2} \Delta t_{1d}$.
We conclude that the extremely large values of the Doppler factors, plus the lack of simultaneous optical and X-ray variability to the GeV and sub-TeV flare, strongly disfavor one-zone models, pointing instead to a two-zone model, as discussed in the case of the flaring phase of the FSRQ PKS 1222+216 \citep{ta11}.

Another important element to consider for the modelling is the huge opacity for gamma rays characterizing the innermost regions of a FSRQ.
In particular, gamma rays with energies exceeding a few tens of GeV produced within the radius of the BLR would be strongly attenuated.
Therefore, the highest energy part of the spectrum, observed by MAGIC, should have been emitted close to or beyond the BLR radius (see e.g. \citealp{pa14} and references therein).

\subsection{Two zone external Compton model}
Considering the conclusions above, we reproduced the broadband emission of the source with a two-zone model, inspired by the scenario c) of \citet{ta11}. 
The two emission regions are moving with the same Doppler factor along the jet. 
We take the simplest assumption that the first emission region is located, as in the case of other FSRQs, inside the BLR.
The opacity condition forces however the second emission region, the production site of the VHE gamma rays, to be outside of the BLR.
The gamma-ray emission is the sum of the SSC and EC components on the radiation field of BLR and dust torus.
Both radiation fields are included in the calculations of both emission zones, however the BLR radiation field dominates the EC in the zone closer to the black hole, and the radiation field of the torus dominates the farther zone. 
The luminosity of the accretion disk is taken to be $L_{\rm d}=6\times 10^{44}$ erg s$^{-1}$ \citep{gtf10}. 
This value is quite low, if compared to a typical FSRQ.
The radius of the BLR and that of the torus, calculated accordingly to the scaling laws of \citet{gt09}, are $R_{BLR}=7.7\times 10^{16}$\,cm and $R_{torus}=2\times 10^{18}$\,cm.

According to the scenario presented here, the GeV and sub-TeV emission is mostly produced in the EC and SSC process in the farther region (see orange curves in Fig.\,\ref{fig:mwlsed}).
This allows it to escape strong absorption of sub-TeV emission in the BLR radiation field. 
The size of the emission region is sufficiently small to account for the one-day variability timescales observed in this energy band (see Fig.\,\ref{fig:lc}). 
On the other hand the optical and X-ray emission comes mostly from the inner region. 
Lack of strong variability in those energy bands seen in Fig.\,\ref{fig:lc} points to the stability of the emission from this region on the timescales of at least a fortnight. 
It is also self-consistent with the procedure of demagnification of the flux described in Section~\ref{sec:swift2}.  
We recall that blazar emission models, reproducing the innermost regions of the jet (distance from the black hole below $1$\,pc), cannot account for the radio emission (frequencies at which the region is optically thick) which, instead, is produced by farther, optically thin regions of the jet. 
The spatial separation of ``Jet in'' and ``Jet out'' might in principle introduce a delay between emission observed from them. 
If the same population of electrons, traversing along the jet, encounters first ``Jet in'' and afterwards ``Jet out'' we can expect to observe a delay of:
$\sim (1+z_s)\Delta R_{dist}/(c\Gamma D)$, where $\Delta R_{dist}$ is the distance between the two regions.
Using the modeling paramaters reported in Table~\ref{tab:param} one would obtain a time delay of only $\sim$6.9 h, which is significantly shorter than the duration of the flare, and very small on the temporal scale of Fig.~\ref{fig:lc}.
Moreover the delay would not be observable if, as assumed above, the emission from the ``Jet in'' region is quasi-stable.

% -------------------------------------------------------------
%TAB: Model
\begin{table*}
\begin{center}  
\begin{tabular}{llllllllllll} 
\hline  
   &$\gamma_{\rm min}$ &$\gamma_{\rm b}$ &$\gamma_{\rm max}$ &$n_1$ &$n_2$ &$B$ [G] &$K$ [cm$^{-3}$] &$R$ [cm] & $R_{dist}$ [cm]
   &$\delta $ &$\Gamma$ \\
\hline
Jet in   &2.5     &300   &$3\times 10^4$ &2    &3.9 &1.1  &$1.5\times 10^5$ &$7\times 10^{15}$ & $7\times 10^{16}$ &20 &17 \\
Jet out &$10^3$ &  $7\times 10^4$ &$2\times 10^5$ &2 &4.3   &0.03    &$3\times 10^7$   &$ 10^{15}$ &$2\times 10^{17}$ &20    &17 \\
\hline\\
\end{tabular}
\end{center}
\vspace{-0.3cm}
\caption{
Input parameters for the emission model shown in Fig.\,\ref{fig:mwlsed} ``Jet in'' and ``Jet out'' indicate the emission regions located inside or outside the BLR respectively. 
The parameters are: the minimum ($\gamma_{\rm min}$), break ($\gamma_{\rm b}$) and maximum  ($\gamma_{\rm max}$) 
Lorentz factor and the low energy ($n_1$) and the high energy ($n_2$) slope of the smoothed power law electron 
energy distribution, the magnetic field $B$, the normalization of the electron distribution, $K$, the radius of 
the emission region, $R$, the distance from the central BH at which the emission occurs, $R_{dist}$, the Doppler factor $\delta$ and the corresponding bulk Lorentz factor $\Gamma$. 
Doppler factors are calculated assuming that the observer lies at an angle $\theta_{\rm v}=2.8^\circ$ from the jet axis. }
\label{tab:param}
\end{table*}
% -------------------------------------------------------------

The parameters for the region inside the BLR  (Table \ref{tab:param}) are in the range of those typically derived for a FSRQ with leptonic models (e.g. \citealp{gt15}). 
For the outer regions there is a strong constraint on the luminosity of the synchrotron component, which - given the large Lorentz factors of the electrons required to produce the high-energy component - peaks in the UV-soft X-ray band. 
To keep the synchrotron component below the limits and, at the same time, reproduce the powerful high-energy IC component, the magnetic field must be kept to quite low values. 
This is similar to the case of PKS 1222+216 discussed in \citet{ta11}. 
As in that case, a possibility to explain such low values could be to assume that this emission region is the product of processes involving magnetic reconnection, in which magnetic energy is efficiently converted to electron energies (e.g. \citealp{si15}).

\section{EBL constraints}\label{sec:ebl}
The VHE gamma-ray observations of distant sources can be used to constrain the level of EBL.
A wide range of methods have been applied in the past, starting from comparing spectral shape in the unabsorbed GeV range with the one in the TeV range (e.g. \citealp{dw94}), and progressing to more elaborate methods, such as testing a grid of generic EBL spectral shapes and excluding the ones resulting in a pile up (i.e. convex spectrum) or a too hard intrinsic spectrum \citep{mr07}.
More recently population studies have been performed.
In particular \citet{ab13} used the specific shape of the EBL-induced feature in the spectrum. 
They performed a joint fit of multiple TeV spectra of sources with redshift $z<0.2$ assuming smoothness of the intrinsic spectrum and putting constraints on the normalization factor of an EBL model with $\sim 15\%$ precision.
A somewhat different approach was applied by \citet{ack12} to \fermilat\ data of about 150 BL Lacs. 
In this case low photon statistics in the sub-TeV regime does not provide a good handle on the particular spectral shape of the feature. 
However, as the observations below a few tens of GeV are not affected by the EBL absorption, the authors could obtain a direct measure of the absorption at the sub-TeV energies, assuming that the intrinsic spectral shape can be described by a log parabola. 
In the last class of methods (see e.g. \citealp{do13}) instead of generic function shapes in TeV, the spectral form based on modelling of broadband emission with synchrotron-self-Compton scenario is used. 

Even though the above methods applied to nearby ($z\lesssim0.2$) sources improved greatly our EBL knowledge in this redshift range, and the detection of 3C~279 \citep{al08} led to a major revision of the EBL models, the measurements at higher redshift are still sparse and are burdened with large uncertainties. 
The $1\sigma$ error band of the \citet{ack12} measurement for the sources with redshift $0.5 < z < 1.6$ allows for about a factor of two uncertainty in optical depth for EBL absorption.
More recently, PKS1441+25 observations with MAGIC and VERITAS resulted in constraints on the scaling factor of optical depth predicted by the current EBL models to be $\lesssim 1.5-1.7$ \citep{ah15, ab15}.

Here we use the \fermilat\ and MAGIC data collected from \srcs\ during the flare to perform an independent measurement of EBL absorption at $z=0.944$.
Since the two instruments measured a similar time scale of the flare it is plausible to assume that the GeV and sub-TeV emission originates from the same emission region.
This assumption is further supported by the SED modelling presented in Section~\ref{sec:sed}.
The dependence of the size of the emission region on the energy, combined with microlensing, might affect the observed GeV spectrum (see Section~\ref{sec:lens}), introducing additional systematic uncertainties in the derived constraints on EBL.
The spectrum observed by MAGIC from \srcs\ gives us a chance to probe the EBL at wavelengths of $\sim 0.3-1.1\mathrm{\,\mu m}$.
We use a method adapted from \citet{ab13}. 
We perform a joint spectral fit combining \fermilat\ and MAGIC points using a set of possible spectral shapes.
To cover better the energy range of the EBL induced cut-off for this study we use a finer binning of the MAGIC data than presented in Fig.\,\ref{fig:magicsed}, resulting in 5 bins. 
The intrinsic spectral shapes are attenuated by EBL according to optical depths presented in \citet{do11}; however we allow an additional scaling parameter $\alpha$ of the optical depth.
The following spectral models (power law, power law with a cut-off, log parabola, log parabola with a cut-off) are used:
\begin{eqnarray}
\mathrm{PWL:\ } dN/dE&=&A E^{-\gamma}, \label{eq3} \\
\mathrm{PWLCut:\ } dN/dE&=&A E^{-\gamma} \exp(-E/E_{cut}), \\
\mathrm{LP:\ } dN/dE&=&A E^{-\gamma -b \log{E}}, \\
\mathrm{LPCut:\ } dN/dE&=&A E^{-\gamma -b \log{E}} \exp(-E/E_{cut}), \label{eq6}
\end{eqnarray}
where we apply additional source physics-driven conditions: $E_{cut}>0$, $b>0$. 
For each spectral shape we compute the $\chi^2$ value of the fit as a function of $\alpha$.
We determine the best fit and the best estimation of $\alpha$ from the minimum of such a curve.
The $1\sigma$ statistical uncertainty bounds of the $\alpha$ parameter can be obtained as the range of $\alpha$ in which the $\chi^2$ increases by $1$ from the minimum value. 

\begin{figure*}
\includegraphics[width=0.33\textwidth]{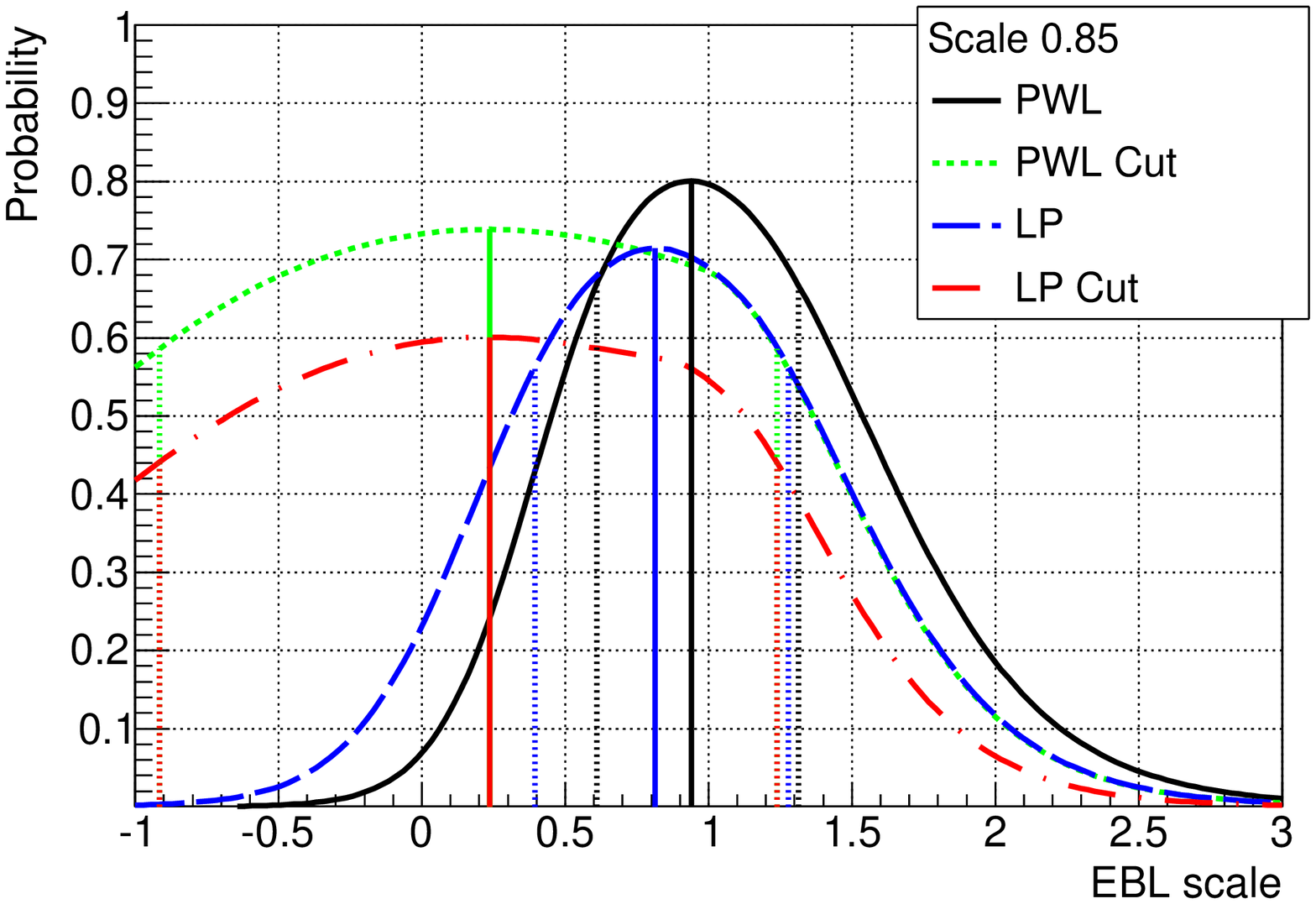}
\includegraphics[width=0.33\textwidth]{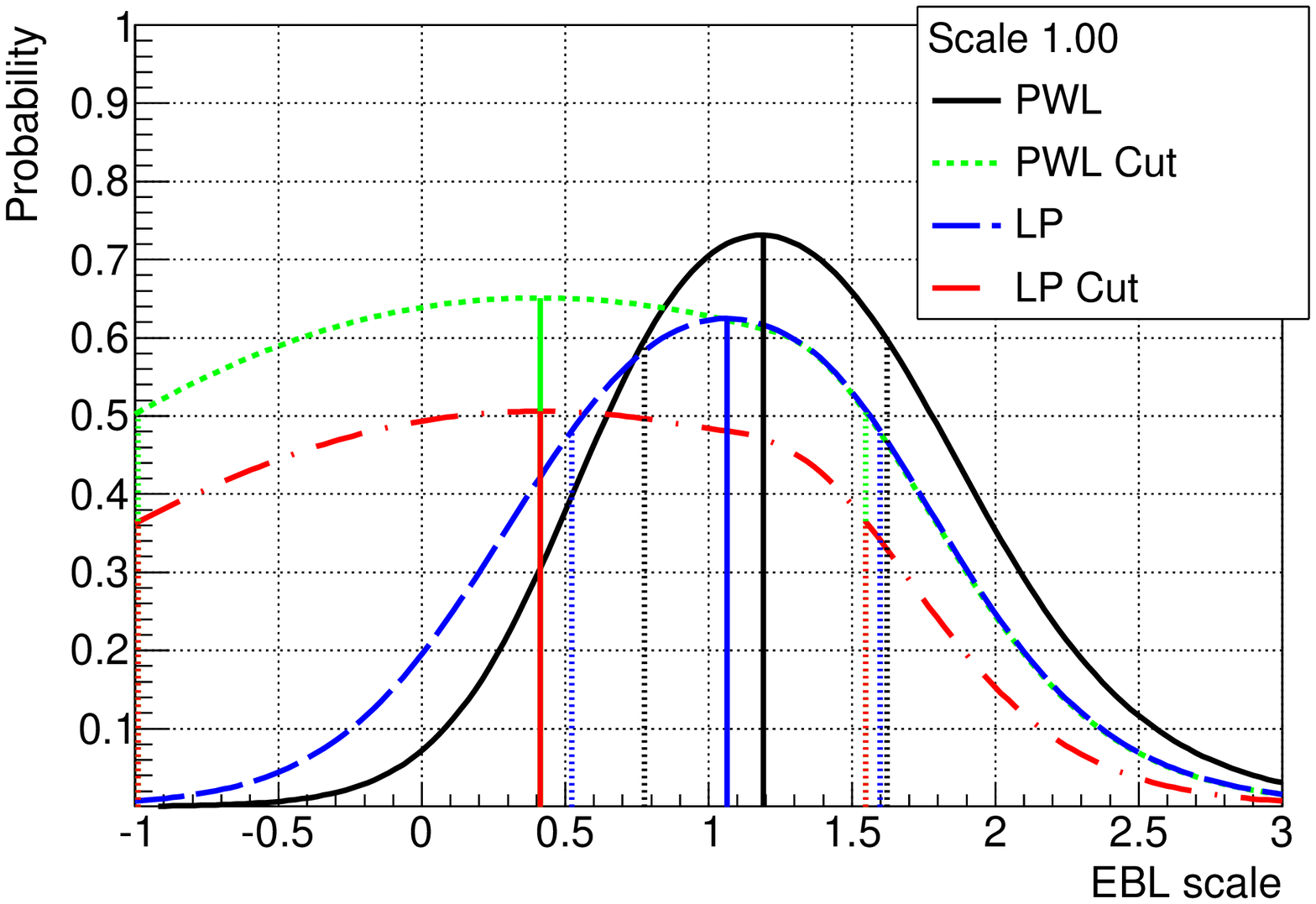} 
\includegraphics[width=0.33\textwidth]{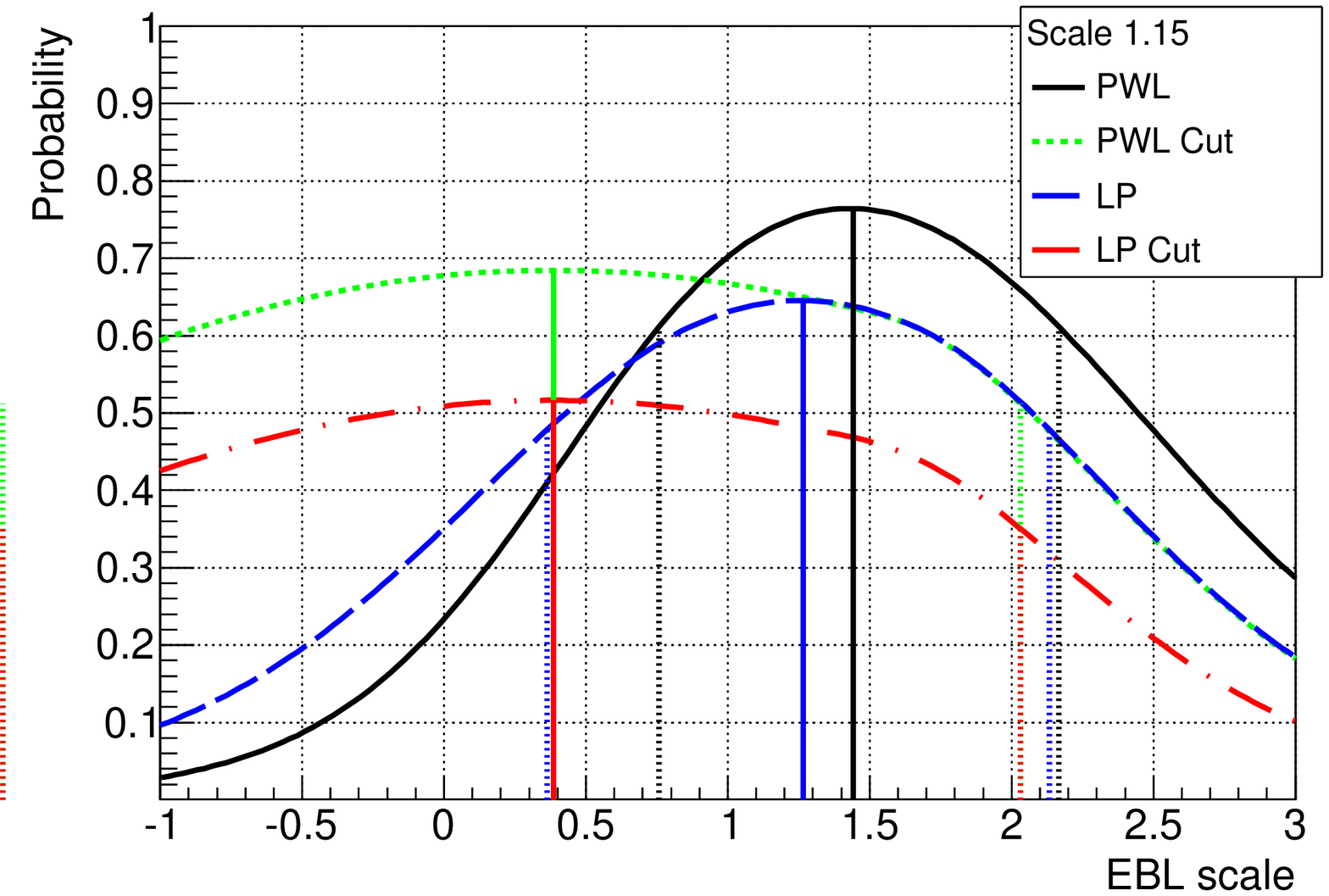}
\caption{
Probability of a SED fit as a function of the EBL scaling parameter.
Different styles and colors of the lines represent different spectral shapes: power law (solid, black), power law with an exponential cut-off (dotted, green), log parabola (long dashed, blue), log parabola with exponential cut-off (dot-dashed, red). 
The vertical lines show the scaling for which the best probability is obtained (solid) and $+1$ change in $\chi^2$ of the fit from this maximum (dotted). 
Nominal light scale of MAGIC is assumed in the middle panel.
The light scale is decreased (increased) by 15\% in left (right) panel. 
}\label{fig:eblprob}
\end{figure*}

The fit probability as a function of the EBL scaling parameter is shown in the middle panel of Fig.\,\ref{fig:eblprob}. 
Out of the phenomenological function shapes (Eq.~\ref{eq3}-\ref{eq6}) the highest fit probability is obtained with the simple power-law spectral model.
Using this spectral model we obtain an estimation of the EBL scaling parameter of $\alpha = 1.19\pm0.42_{\rm stat}$ at a redshift of $0.944$.
Such an assumption of a single power-law between 3 and 200 GeV, even though slightly preferred by the best fit probability, might be at odds with the FSRQ emission models. 
The spectral models allowing for an intrinsic curvature/cutoff exhibit a slower dependence of $\chi^2$ on the EBL scaling for the low values of $\alpha$ resulting in less constraining bounds. 
Notably, all the tested spectral shapes provide a $1\sigma$ upper bound below the value for a simple power-law spectral shape.
Spectral shapes with an additional intrinsic cut-off result in only a small increase of $\chi^2$ from the scaling factor of 1 (nominal EBL) to 0 (no EBL).
Therefore no strict lower bound can be derived on $\alpha$. 

Systematic uncertainties can affect the obtained results. 
In particular, since we use a combination of MAGIC and \fermilat\ data, a shift in energy scale or flux normalization between the spectra obtained from the two experiments could affect the result. 
Due to a very steep spectrum in the sub-TeV range the dominant systematic effect is the 15\% uncertainty of the energy scale of MAGIC \citep{al16b}.
It can shift the reconstructed spectrum in energy causing a large, up to 40\%, shift in estimated flux. 
This effect is much larger than the pure flux normalization uncertainty reported in \cite{al16b}.
The uncertainty of the spectral slope, due to the limited energy range of the spectrum, has a negligible effect. 
Finally the systematic uncertainty of \fermilat\ (see e.g. \citealp{ack12b}, where 2\% accuracy on the energy scale is reported) is also negligible compared to the ones of MAGIC in the case of this source. 
Therefore in order to investigate the systematic uncertainty on the EBL scaling parameter we performed full analysis using telescope response with a modified light scale by $\pm15\%$ following the approach in \cite{al16b} and \cite{ah16}.

In all three cases (see Fig.\,\ref{fig:eblprob}) the best probability of the fit out of the assumed phenomenological function shapes is obtained with a simple power-law fit. 
However, the corresponding EBL scaling parameter shifts by $0.25$ for a power-law case.
Also the statistical error for an increased light scale in this case is slightly larger.
Therefore, allowing for an intrinsic curvature (log-parabola spectral shape, and/or an exponential cut-off) we obtain a 95\% C.L. upper limit of $\alpha < 2.7$.
This limit is less constraining than the one obtained with PKS 1441+25 \citep{ah15}.

We repeated the analysis substituting the EBL model of \cite{do11} by other currently considered models: \cite{fr08}, \cite{fi10}, \cite{gi12}, \cite{in13}.
In the case of all the models the highest fit probability was obtained with a power law spectrum. 
The results of the best scaling parameter of the optical depth of these models are summarized in Table~\ref{tab:eblmodels}.
\begin{table}[t]
\begin{center}
\begin{tabular}{l|l|l}
%Model     & Scale (PWL) & Scale(all)  \\ \hline\hline
Model     & $\alpha$ (PWL) & $\alpha$ (all)  \\ \hline\hline 
\cite{fr08} & $1.19 \pm 0.42_{\rm stat} \pm 0.25_{\rm syst}$ & $<2.8$ \rule{0pt}{12pt}\\[0.5ex]\hline
\cite{fi10} & $0.91 \pm 0.32_{\rm stat} \pm 0.19_{\rm syst}$ & $<2.1$ \rule{0pt}{12pt}\\[0.5ex]\hline
\cite{do11} & $1.19 \pm 0.42_{\rm stat} \pm 0.25_{\rm syst}$ & $<2.7$ \rule{0pt}{12pt}\\[0.5ex]\hline
\cite{gi12} & $0.99 \pm 0.34_{\rm stat} \,_{-0.18\, \rm syst} ^{+0.15\, \rm syst}$ &$<2.1$ \rule{0pt}{12pt}\\[0.5ex]\hline
\cite{in13} & $1.17 \pm 0.37_{\rm stat} \,_{-0.13\, \rm syst} ^{+0.10\, \rm syst}$ &$<2.2$ \rule{0pt}{12pt}\\[0.5ex]
\end{tabular}
\caption{
Limits on the scaling parameter $\alpha$ of the optical depths in various EBL models.
The second column specifies the limit for the intrinsic spectral model with the highest peak probability from the assumed phenomenological spectral shapes (Eq.~\ref{eq3}-\ref{eq6}). 
For all the EBL models it is the power law shape. 
The last column specifies the 95\% C.L. limit allowing all considered spectral shapes and 15\% energy scale systematic uncertainty. 
}
\label{tab:eblmodels}
\end{center}
\end{table}
As in the case of the model of \citet{do11} we report the limits on the optical depth scaling factor for a power-law intrinsic spectral shape and a more conservative 95\% C.L. upper limit for intrinsic spectral shapes allowing an arbitrary steepening or a cut-off.  
The combined \fermilat\ and MAGIC spectrum is consistent with all five EBL models considered here.

%
%______________________________________________________________
\section{Conclusions}
MAGIC has detected VHE gamma-ray emission from \srcs\ during the trailing component of a flare in July 2014. 
It is currently the most distant source detected with a ground-based gamma-ray telescope, and the only gravitationally lensed source detected in VHE gamma-rays.
The VHE gamma-ray emission lasted for two nights achieving the observed flux of $\sim 30\%$ of Crab Nebula at 100\,GeV.
Using the EBL model from \citet{do11}, the intrinsic spectral index in this energy range was found to be $2.35\pm0.75_{\rm stat} \pm 0.20_{\rm syst}$.
The VHE gamma-ray flare was not accompanied by a simultaneous flux increase in the optical or X-ray energy range. 
We have modeled the X-ray emission as a sum of two components with different magnifications, the weaker one absorbed with column density of $(2.4\pm0.5)\times 10^{22}\ \mathrm{at.\,cm^{-2}}$.
The combined \fermilat\ and MAGIC energy spectrum is consistent with the current EBL models.
These constraints are however not very strong, with the EBL density scaling parameter being less than 2.1-2.8 of the one predicted by the tested models. 
The broadband emission of \srcs\ is modeled in a framework of a two-zone external Compton model. 
According to this scenario, the quasi-stable optical and X-ray emission originates mostly in the inner zone. 
The enhanced gamma-ray emission during the flare is produced in the second zone, located outside of the BLR.

\begin{acknowledgements}
We would like to thank
the Instituto de Astrof\'{\i}sica de Canarias
for the excellent working conditions
at the Observatorio del Roque de los Muchachos in La Palma.
The financial support of the German BMBF and MPG,
the Italian INFN and INAF,
the Swiss National Fund SNF,
the ERDF under the Spanish MINECO (FPA2012-39502), and
the Japanese JSPS and MEXT
is gratefully acknowledged.
This work was also supported
by the Centro de Excelencia Severo Ochoa SEV-2012-0234, CPAN CSD2007-00042, and MultiDark CSD2009-00064 projects of the Spanish Consolider-Ingenio 2010 programme,
by grant 268740 of the Academy of Finland,
by the Croatian Science Foundation (HrZZ) Project 09/176 and the University of Rijeka Project 13.12.1.3.02,
by the DFG Collaborative Research Centers SFB823/C4 and SFB876/C3,
and by the Polish MNiSzW grant 745/N-HESS-MAGIC/2010/0.
\\
%% Fermi acknowledgements
The \textit{Fermi} LAT Collaboration acknowledges generous ongoing support
from a number of agencies and institutes that have supported both the
development and the operation of the LAT as well as scientific data analysis.
These include the National Aeronautics and Space Administration and the
Department of Energy in the United States, the Commissariat \`a l'Energie Atomique
and the Centre National de la Recherche Scientifique / Institut National de Physique
Nucl\'eaire et de Physique des Particules in France, the Agenzia Spaziale Italiana
and the Istituto Nazionale di Fisica Nucleare in Italy, the Ministry of Education,
Culture, Sports, Science and Technology (MEXT), High Energy Accelerator Research
Organization (KEK) and Japan Aerospace Exploration Agency (JAXA) in Japan, and
the K.~A.~Wallenberg Foundation, the Swedish Research Council and the
Swedish National Space Board in Sweden.

Additional support for science analysis during the operations phase is gratefully acknowledged from the Istituto Nazionale di Astrofisica in Italy and the Centre National d'\'Etudes Spatiales in France.

\end{acknowledgements}

%-------------------------------------------------------------------

\end{document}